\documentclass[12pt]{article}
\usepackage[dvips]{graphicx}

\newcommand{\ket}[1]{\,\left|\,{#1}\right\rangle}

\begin {document}
\begin{flushright}
{\small
SLAC--PUB--8964\\
October 2001\\}
\end{flushright}

\begin{center}
{\bf\LARGE \boldmath The $CP$ Puzzle \unboldmath in the Strong
Interactions}\footnote{Work supported by Department of Energy
contract DE--AC03--76SF00515.}

\vfill

\bigskip
{\it Helen Quinn \\
Stanford Linear Accelerator Center \\
Stanford University, Stanford, California 94309 \\
E-mail:  quinn@slac.stanford.edu}
\medskip
\end{center}

\vfill

\begin{center}
{\bf\large
Abstract }
\end{center}

This lecture, directed to an broad audience including non-specialists,
presents a short review of the problem of strong $CP$ symmetry
maintenance.  The problem is defined and the possible solutions briefly
reviewed.  I discuss the way in which Roberto Peccei and I came up with
one solution, generally known as Peccei-Quinn symmetry.  \vfill

\begin{center}
{\it
Dirac Medal Lecture 2001\\
Presented at the International Centre for Theoretical Physics \\
Trieste, Italy\\
July 3, 2001 }\\
\end{center}

\vfill \newpage

I would like to begin by thanking the ICTP for this honor, which I
greatly appreciate receiving.  To Miguel Virasoro and to the rest of the
selection committee I want to express my sincere gratitude.  I am
greatly honored to be invited to join the company of the distinguished
physicists who have received this award in the past.  I also want to
thank my collaborators on the work cited in this award, Steve Weinberg
and my co-recipient Howard Georgi on the unification of the
couplings \cite{hierarchy} and Roberto Peccei, with whom I did the work I
will talk about today \cite{pqsymmetry}. He was not included in this
award as it was principally focused on the other work, on Grand Unified
Theories.  I know that Jogesh Pati, the third co-winner this year,has
talked on that topic, not only in his Dirac lecture \cite{patidirac} but
also in his lectures for the Particle Physics school \cite{patilectures}.
So I have chosen to devote my talk today to the topic of
strong $CP$ Violation and how to avoid it.  This lecture thus also is an extension
of the course on $CP$ violation that I am giving in the school here this
week \cite{quinnlectures}. This lecture is intended for a broader audience
than usual for such a topic, so I begin by an attempt to explain the topic
to the less technically-expert part of the audience.

The  term $CP$
is the technical name of a symmetry that appears to be almost but
not quite an exact symmetry of nature, namely the symmetry between
the laws of physics for matter and those for antimatter
\cite{pub8784}. We know from experiment that this symmetry applies
to very high precision in the strong interactions.  These are the
interactions responsible for binding quarks together to make the
observed particles such as protons and neutrons, and also for the
forces between such particles, for example those that cause the
protons and neutrons to bind together in atomic nuclei.  We also
find that this symmetry applies for electromagnetic interactions
(those due to electric and magnetic charges and fields). The
surprise of the 1960's was that it is not quite true for the weak
interactions.  Weak interactions are those responsible for many
types of particle decays, in particular all those where one type
of quark changes to a different type.  The surprise was the observation of
weak decay processes that would be forbidden if the matter-antimatter
symmetry were exact, rare decays of the long-lived neutral kaon, which should be
a $CP$ odd state, to the $CP$-even state consisting of two pions \cite{CCFT}.

In physics exact symmetries are easy to include in a theory.
Equally easily we can write theories that do not have the symmetry at
all.  The hardest situation to explain is one where we seem to
have an almost exact symmetry.  Typically once a symmetry is broken
there is no reason for it to appear to be close to true, rather it just
disappears altogether.  So when we find an almost true symmetry we need
to find a mechanism that can explain that property.  In most cases it is
no problem to have different symmetry structure for the different
interactions.  Indeed that is really what distinguishes one type of
interaction from another in our theory.  So one  might think that all we had to do was
explain why there is a small $CP$ violation in weak decays.
That can be readily accommodated
in the three-generation Standard Model.

It turns out however that for the particular case of matter-antimatter symmetry,
or $CP$ this separation in the symmetry properties of the different interactions is
not so readily achieved. In our current Standard Model theory (and in any extension of
it which maintains the well-established QCD theory for the strong interactions) any
breaking of $CP$ symmetry in the weak interactions can, and generally will, induce a
breaking of that symmetry in the strong interactions. So the challenge is to find a
class of theories where this is not the case, or at least where the
magnitude of the effect can be tightly controlled.

\begin{figure}
\centering
\includegraphics{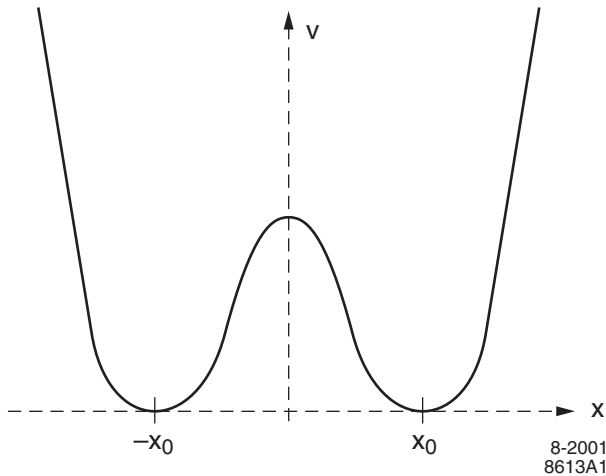}
\caption[*]{Potential well with two degenerate minima.}
\label{fig:1}
\end{figure}

In order to describe how this strong $CP$-breaking comes about, and
then how it can be avoided, I will begin by describing some
similar quantum mechanical properties for a much simpler system. The
features I will stress in this more familiar problem all have parallels
in the strong $CP$ problem. Consider a
particle in a (one-dimensional) potential well such as that shown
in Fig.~\ref{fig:1}. We want to examine the ground states of a particle in this well.
The potential is symmetric about the line
x=0, but rather than having a single lowest-energy point at that
location it has two equally low-energy locations at $x=\pm x_0$.
If this were a classical physics problem, say a ball rolling on a
hilly surface, and you were asked to find the lowest energy state
for the ball, you could readily see that there would be two equally low energy states,
one centered at each of the two minima.  Indeed, even in quantum mechanics,
by making an approximation that each well is independent, one can
find two such states, which are described by probability
distributions centered on either minimum.  Let us call these
states $\psi_{\rm Left}$ and $\psi_{\rm Right}$.  But the quantum phenomenon of
tunneling allows a particle located at some time at one minimum
to have a finite probability to appear at the other minimum some time later.
The two wells are not truly separate. The states
$\psi_{\rm Left}$ and $\psi_{\rm Right}$ cannot be the true stable states
(eigenstates) of the system.

To find the correct stable lowest-energy states we apply a general rule
of quantum physics.  Any symmetry of the energy function, the
Hamiltonian, must also be manifest in the stable states of the system.
all such states can be labeled by the way in which they transform under
the symmetry.  The potential energy has a symmetry under the
transformation of $x$ to $-x$.  This is called a Parity transformation,
it transforms left into right and vice-versa.  (In a three-dimensional
space parity reverses all directions).  The rest of the energy, the
kinetic energy, is clearly also invariant under this transformation,
since it proportional to is the square of velocity.  This means that the
entire Hamiltonian for our problem has an invariance under the Parity
transformation.  Quantum physics then tells us that all stable physical
states must also be transformed into themselves under Parity.  Since two
reflections is equivalent to no reflection at all, the only consistent
choices for the constant coefficients, or quantum numbers, of the states
under the Parity operation are $\pm 1$.  The superpositions $[\psi_{\rm
Left}\pm \psi_{\rm Right}]/\sqrt{2}$ are the stable ground states with
these properties.

As previously stated, the states $\psi_{\rm Left}$ and $\psi_{\rm
Right}$ are broad probability distributions rather than the simple point
locations of the classical solutions.  Each distribution has a small
tail that reflects the fact that there is some small probability to find
the particle in the other minimum.  Thus the two distributions overlap
slightly.  Hence they interfere with one another.  In one of the
definite-Parity states the interference adds a little to the energy, and
in the other it reduces it slightly.  There is a unique lowest energy
state, which turns out to be the combination $[\psi_{\rm Left} +
\psi_{\rm Right}]/\sqrt{2}$.  This state transforms into exactly the
same state under parity, with coefficient $+1$.  The lowest state which
goes into itself times $-1$ under the Parity transformation, $[\psi_{\rm
Left} -\psi_{\rm Right}]/\sqrt{2}$, has a slightly higher energy.  We
call this an odd-Parity state.  Not only do the left and right states
get mixed up because of tunneling, but also the two different Parity
admixtures have slightly different energy.

 All the higher-energy states of the system can likewise be divided into those
that are even and those  that are odd under the Parity
transformation.  If we add electromagnetic interactions in our theory and
investigate the possibility of radiative
transitions between states of different energies,
we find that all such transitions occur only between states
that have the same Parity.  The Hilbert space, the set of all states of the system,
is split into two disjoint parts, which act as if they are two separate worlds,
knowing nothing of one another. All this is quite familiar
to anyone who has taken a quantum mechanics course. I review it
here because the story about strong $CP$ violation concerns a very similar
phenomenon, but in a less familiar context.

QCD, the theory of the strong interactions, is a gauge theory.
This name labels a class of theories where there are a set of
locally-defined redefinitions of all fields, known as gauge
transformations, that change the form of the fields everywhere but
do not change the energy associated with these fields.  This
invariance, like the Parity invariance in the example above, leads
to a potential that does not have a unique minimum.  In the case
of QCD, because of the non-Abelian nature of the algebra of gauge
transformations it turns out that there are not just two minima to
the potential energy, but an infinite number.  Any state, that is
any static configuration of fields that can be defined by making a
time-independent gauge-transformation of the configuration with
all QCD fields equal to zero has the same energy, namely zero, as
does the state with no fields at all.  We call these pure-gauge
field configurations.  We also require that the fields vanish at
spatial infinity.  (I here use the language of states, which I
describe by static gauge field configurations, to make this story
a little more understandable,I hope . I may gloss over some of the
finer distinctions between Hamiltonian quantum mechanics and
Euclidean field theory in this tale.  I do not attempt to define
what I really mean by states in a continuum field theory, rather I
am trying to give you an intuitive picture how the strong
$CP$-violating term in the theory arises.  One usually sees this
problem discussed in the Eucliean Field theory, I find the
Hamiltonian discussion more intuitive, so that is why I present
that language.  If you want to see this language are discussed in
some detail for a similar situation read my paper with Marvin
Weinstein on the two dimensional Abelian Higgs theory
\cite{onedscalar}.)

Among the possible gauge transformations there are some for which the
fields so generated cannot be deformed back to the zero-field case by
changing them smoothly.  For any static field configuration one can
compute the the quantity \begin{equation} n= (1/32 \pi^2)\int d^4x
\epsilon^{\mu\nu\rho\sigma}F_{\mu\nu}^a F_{\rho\sigma}^a \ .
\end{equation} This is a toplogical quantity which must be an integer
for any pure gauge field, it is called the winding number
\cite{thooft}. The quantity
$(1/2)\epsilon^{\mu\nu\rho\sigma}F_{\mu\nu}^a F_{\rho\sigma}^a $
can also be seen to be a total derivative of the quantity
\begin{equation}
K^\mu =\epsilon^{\mu\nu\rho\sigma}[A_\nu^a
F_{\rho\sigma}^a + f^{abc}A_\nu^aA_\rho^bA_\sigma^c].
\end{equation}
(For more details of this story see for example the TASI lectures
on the strong $CP$ problem given by Michael Dine \cite{tasi}.) The
winding number can thus be written as a surface integral over
$(1/16\pi^2)K_\mu$.  For fields which vanish at spatial infinity
this integer must in fact be the difference between two integers,
integers that label the state at time minus infinity and time plus
infinity.  So our static gauge field configurations can be
labelled by integers n, which are the spatial integral of
$(1/16\pi^2)K_0$.  Any gauge field where this quantity is 1 cannot
be continously transformed to one where it is zero, since that is
a discontinuous jump.  To get smoothly such a configuration with
label 1 to one with label 0 we must pass through some field
configurations that are not pure-gauge fields and that therefore
have higher energy than a pure gauge field.

\begin{figure}[h]
\centering
\includegraphics{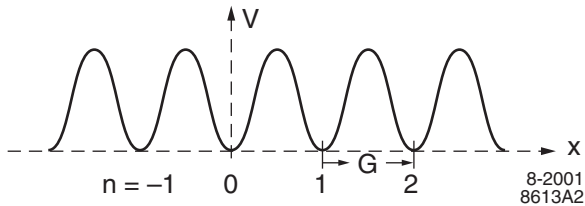}
\caption[*]{Cartoon of the potential for a non-Abelian Gauge
Theory.} \label{fig:2}
\end{figure}

So now we can make a cartoon of this situation, shown as
Fig.~\ref{fig:2}. We plot the energy of the gauge-field
configurations as the vertical axis and the horizontal axis
labeled $A$ schematically represents all possible field
configurations over all of space.  The only meaningful part of
this picture is that there are now a series of minima to the
potential.  We can label these minima by the integers $n$, (the
spatial integral of $(1/16\pi^2)K_0$ for the pure gauge field
configuration corresponding to such a minimum).  We denote these
pure-gauge states by $\ket n$.  There is a well-defined gauge
transformation, which I will call $G$, which transforms the state
with field configuration $\ket n$ into $\ket{|n+1}$ for any $n$.
In the two-well case we saw a tunneling probability between the
different minima of the potential, likewise here there is a
tunneling possibility between states of different $n$. Such a
tunneling event is called an instanton.  (More precisely an
instanton is a classical solution of the Euclidean field theory
that has winding number 1.)

Just as quantum mechanics told us we must find states of definite Parity
for the double well potential, the rules of quantum physics here say
that the physical ground states must be those that map into themselves,
times a constant, under the non-trivial gauge transformation G. Such
states can easily found, they are the superpositions
\begin{equation}
\ket \theta = \Sigma_n e^{-in\theta}\ket n \ .
\end{equation} These are
an infinite set of such states, one for each choice of $\theta$.
Such a state is not changed by $G$, except that it is multiplied
by a constant or quantum number:
\begin{equation}
G\ket \theta = e^{i\theta}\ket \theta\ .
\end{equation}
In the
problem where the potential had two minima there were two possible
ground states, here with infinitely many degenerate minima of the
potential there are infinitely many possible ground states. States
of different theta are the physically distinct vacua for the
theory, each with a distinct world of physics built upon it, just
as the Parity odd and Parity even worlds of the two-well potential
are distinct.  (Translated back into the Euclidean field theory
this discussion tells us there are an infinite set of theories,
those with a term of the form of equation 1 added to the
Lagrangian with coefficient $\theta$ \cite{thooft}.)

All this may seem to be just formal manipulation, but a little
examination shows that $CP$ conservation in the strong
interactions is only exactly true in the world built on the state
$\theta=0$.  The integrand in Equation 1 is a $CP$-violating
quantity.  Thus the $\theta$-dependent term in the action induces
$CP$-violating effects. Conversely, any $CP$-violating
interaction, even in the weak interaction sector of the theory,
will, in general, induce a non-zero value of $\theta$ via
loop-effects.  When we investigate the consequences of living in a
world with non-zero theta we immediately find this gives a
problem.  The electric dipole moment of the neutron is constrained
by experiments to be extremely small.  But the theory predicts a
value proportional to $\theta$.  This constrains our world to be
one in which theta is of order $10^{-12}$ or smaller
\cite{edmcalc}.  This is then the strong $CP$ puzzle:  why is this
parameter $\theta$ so tiny?  {\em A priori} it could have any
value, from $-\pi$ to $\pi$.  What properties of the theory can
ensure such a small value?

There are, to the best of our current knowledge, three possible answers
to this question.  One of them is the result of work
 that was included in my citation for this award, the work
that I did with Roberto Peccei.  I will
briefly summarize all three, and then give a more detailed story of
the Peccei-Quinn mechanism and the way in which we found it.  We do not
yet know which, if any, of these answers are correct, just as we do not
yet know whether Grand Unified Theories describe nature.  One bold thing
that the Dirac Medal committee has chosen to do is to honor work which
may in the end turn out not to be the correct theory.
However it has stood the test of 25 years of examination and experimental
probing, it is still a possible answer to the problem.

Before we can discuss the three solutions there is one more
feature of the theory that we must discuss, and that is the role of
quark fields in the $\theta$ determination.  If one transforms a
quark field by a factor $e^{i\gamma_5\alpha}$ it turns out that
this transformation induces a shift in the value of theta by an
amount $-\alpha$.  But such a transformation also changes the
phase of the mass term for that quark by an amount $\alpha$.  The
quantity that is unchanged by such transformations is the
difference $\theta_{\rm effective} = \theta -{tr\  \ell n \det} M$
where M is the matrix of all quark masses.  This is the actual
(physically meaningful) $CP$-violating quantity in the theory.  In
my lecture yesterday I stressed that one must examine all possible
phase redefinitions in order to know what differences of phases
are the physically meaningful $CP$-violating phases; this is another
example of that rule.

The result immediately suggests the first and perhaps the simplest
solution to the problem.  In the Standard Model, if there is any
one quark with zero bare mass then a chiral transformation of this
quark field can be used to set $\theta_{\rm effective}$ to zero
with no other consequences.  Thus, in the presence of a massless
quark $\theta$ is not a physically meaningful parameter.  The
question is then whether nature chose this solution.  The lightest
quark is the $u$ quark.  Its mass could simply be a
renormalization effect, that is to say the bare up-quark mass
could be zero and the measured mass is that induced via weak
interaction loop diagrams due to the down-type quark masses.  This
does not seem to give a large enough up quark mass to fit the
observed values \cite{massreview}, but it is just possible that
the effect could be big enough in some extensions of the Standard
Model.  Higher loop effects in any such theory will also induce an
effective theta parameter, which may be sufficiently small.  I
think it is very unlikely that the final solution of the problem
will involve a massless quark, but it is not completely ruled out
with current understanding of quark masses.

The second choice is to impose $CP$ as a symmetry of the full
Lagrangian, thereby setting theta to zero at the renormalization
scale where the theory is first defined.  However since there is
an observed $CP$ violation in nature, albeit in the weak
interaction sector, such a theory must be constructed so $CP$
breaking occurs spontaneously, via soft operators which acquire
$CP$-violating vacuum expectation values. There will then be a
non-zero theta parameter induced by loop effects.  If the theory
is constructed to suppress these effects at the one and two loop
level then the resulting theta parameter can be small enough.
Indeed in the Standard Model the theta parameter only receives
renormalization corrections at the three loop-level. A number of
examples of theories of this type have been suggested in the
literature \cite{softbreakingmodels}.  This answer remains a
viable one.

I want to talk about the third approach in a little more detail, it is
the one devised by me together with Roberto Peccei.  Remember that quark
masses arise in the Standard Model because the Higgs field has a
non-zero vacuum value.  Roberto and I saw that one could add an
additional symmetry to the theory in such a way that it is automatic
that the vacuum energy is minimized for $\theta_{\rm effective} =0$.
Technically this new global U(1) symmetry is not quite an exact
symmetry.  Like the strong $CP$ symmetry itself, it is a
pseudo-symmetry, broken only by non-perturbative or instanton
(tunneling) effects.  This is exactly why it works as desired.  The
trick is to make the Higgs field energy depend on the $\theta$ value in
such a way that, for any initial value of $\theta$, the Higgs field will
choose a vacuum value such that the resulting physical parameter
$\theta_{\rm effective}$ is zero.  The Higgs vacuum expectation values
aquire phases such that the phase of the quark mass matrix cancels
against the initial $\theta$.

The way this idea occurred to us was very much a consequence of
the first solution, the fact that the theta-parameter is
irrelevant for zero quark mass.  In the Standard Model quark
masses are indeed zero in the early Universe, before the phase
transition in which the Higgs field obtains its vacuum expectation
value.  This greatly puzzled me.  How could the theta parameter be
irrelevant in one phase but become relevant in another?  The
answer is that this statement is not quite true; in a general
theory with Higgs fields a chiral redefinition of the quark fields
such as that described above also changes the phases of certain
Yukawa couplings.  Now the question of whether the theta parameter
is physical or not looks a lot like the usual rephasing-invariance
question.  This became clear to me during a conversation in which
Steve Weinberg explained the usual issues of rephasing invariance
and $CP$ violation to me, a conversation which took place while
Roberto and I were still completely mystified by the QCD
$\theta$-dependence.

We know that the quark-Higgs Yukawa couplings are the source of quark masses
once the Higgs field gets a non-zero vacuum value.
This suggested to me the notion that it should be possible to
design a Higgs potential, and choose Yukawa couplings, such that,
no matter what the initial theta value, one would get $\theta_{\rm
effective}=0$ as the Universe cooled, once the quark masses were
induced by the Higgs vacuum expectation value. Roberto and I soon
found models where this was so.  When I described this trick to
Sidney Coleman he pointed out that what we had done was to add to the theory a
U(1) symmetry broken only by instanton effects.  I agreed that was
indeed what we had done, and that is the way we presented it in
our paper.

Let me describe a simple extension of the Standard Model to illustrate
how this idea works.  This simple theory is already ruled out by
experiment, but the generic idea survives.  In the simplest version of
the Standard Model there is a single Higgs weak-doublet field that gives
mass to both the up-type and the down-type quarks.  If we introduce an
additional U(1) symmetry (now called PQ symmetry) under which the
right-handed up and down type quarks transform differently then, to
maintain this symmetry, we must add a second Higgs doublet.  The two
Higgs doublets also transform differently under the PQ symmetry.  One of
them has the Yukawa couplings which give mass for the up-type quarks and
the other for the down-type.  The symmetry also forbids terms of the
type $\phi_1 \phi_2^*$ (and higher powers of this quantity) in the Higgs
potential energy.  However the effects of QCD-instantons provide an
additional contribution to the Higgs potential that violates this rule,
inducing just such a term with a ($\theta$)-dependent coefficient.  The
minimum of the potential then correlates the phases of the quark masses
with $\theta$ in just the way required so that the chiral rotations that
make all quark masses real are exactly those that cancel the initial
$\theta$ value.

The additional pseudo-symmetry has a consequence, as was pointed
out by Weinberg \cite{weinbergaxion} and Wilczek
\cite{wilczekaxion}, namely that there is an additional
pseudo-Goldstone boson, now known as the axion, associated with
it.  The fact that Roberto and I did not notice this obvious
phenomenological consequence of our model shows that we were
focused on the general solution to the strong $CP$ problem.  I, at
least, was so happy to find a general solution to that that I did
not stop to examine other phenomenological implications of the
model we built to demonstrate the idea before we published it.
But the axion implication is common to all such models, for it
arises from the symmetry itself.  Steve Weinberg called me when he
noticed that the theory had an almost zero-mass particle.  He
wanted to ascertain whether we knew it was there.  Our
conversation, as I remember it, went something like this:  Steve
asked whether I had noticed that the U(1) symmetry implies a
pseudo-Goldstone boson.  I saw immediately that he was right, but
told him indeed we had not noticed it.  Then Steve told me that he
too had not at first noticed the obvious symmetry argument but had
found the zero mass eigenstate the hard way, by calculating the
Higgs spectrum in the theory.  He wanted to check whether we
already knew about it before he wrote his paper.  Frank Wilczek
noticed the same effect independently.  By the time the papers
were written Weinberg and Wilczek both used the name ``axion'' for
this particle.  I rather liked the alternate name ``higglet''
which was floating around for a while.

Constraints on the existence of such particles rule out the simple
model described above.  It was quickly eliminated by existing data
and further direct laboratory searches for the predicted axion.
Models which so far elude all constraints have been suggested, the
so-called invisible axion models   \cite{invisibleaxion}.  The
constraints are of three types, from direct laboratory searches,
from astrophysics, and from cosmology.  The initial laboratory
searches looked for the interactions of a penetrating particle in
a detector placed some distance behind the beam dump of an
accelerator \cite{beamdumps}.  More recent searches assume that
the axion is a principle component of the dark matter in our
galactic halo and try to detect the conversion of such a particle
to a photon in an intense electromagnetic field set up in a
carefully-tuned resonant cavity \cite{llnlsearch}.

Astrophysical constraints on light weakly interacting particles
such as an axion arise chiefly from the fact that such a particle
would provide an additional mechanism for energy transport from
the interior of a star to its surface, and hence additional
cooling of the star's core. Constraints of this type can be made
from observation of various astrophysical objects, for example
from the life-time of red giant stars \cite{astroconstraints}.
Astrophysical constraints also come from the observations of
neutrinos produced by the Supernova event known as SN1987A
\cite{sn1987a}.  Although only a few neutrino events were observed
they provide quite stringent restrictions on changes to the model
of such supernova explosions.  The number of neutrinos seen, the
duration of the signal, and its timing relative to the optical
observation of the event were in good agreement with models.  Any
additional particle type that could carry off large amounts of
energy in the early stages of this explosion could strongly alter
the predictions. This puts a bound on the axion parameters.

Cosmological constraints come from the fact that axions are produced
from random Higgs-field fluctuations in the early Universe and survive
as dark matter.  The constraint that axionic dark matter must not
overclose the Universe limits the allowed range of axion parameters.
This constraint is interesting because it acts as an upper bound on the
axion mass, while all other constraints provide only lower bounds.  We
are left with only a relatively small window in parameter space for the
axion.  Recent results suggest that the dark matter density is probably
only about 1/3 of closure density \cite{dasietc} which will further
narrow the available parameter-space for axionic dark matter.

I find it fascinating that an idea to solve a particle physics
problem, that of the small value of the strong interaction $CP$-violating
$\theta$ parameter, should predict a particle of possible cosmological
and astrophysical relevance.  This is a beautiful example of the unity and
universality of physics.  In particle physics we try to understand the
physics of the smallest things, seeking for the basic constituents of
matter and their interactions.  But once we postulate anything at this
level it has consequences. We must examine whether our theory survives
all possible types of constraints. Since astrophysical objects and the
early Universe provide more extreme environments than even our highest
energy accelerators can produce, we must often look to these for
evidence of effects that we cannot directly observe. Eventually we may
even find an axion by a laboratory search based on its role as a
constituent of the dark matter clustered in the halo of our galaxy,
a search that combines cosmology, astrophysics and earth-bound
laboratory physics in a most beautiful way.  A positive result would
certainly be exciting!

\end{document}